# Radiative Thermal Diode Mediated by Nonreciprocal Graphene Plasmons Waveguides

Yong Zhang[1,2], Cheng-Long Zhou[1,2], Hong-Liang Yi[1,2,*], and He-Ping Tan[1,2]

[1]*School of Energy Science and Engineering, Harbin Institute of Technology, Harbin 150001, P. R. China*
[2]*Key Laboratory of Aerospace Thermophysics, Ministry of Industry and Information Technology, Harbin 150001, P. R. China*

**ABSTRACT**

A thermal diode based on the asymmetric radiative heat transfer between nanoparticles assisted by the nonreciprocal graphene plasmons waveguides is proposed in this work. The thermal diode system consists of two particles and a drift-biased suspended graphene sheet in close proximity of them. Nonreciprocal graphene plasmons are induced by the drift currents in the graphene sheet, and then couple to the waves emitted by the particles in near-field regime. Based on the asymmetry with respect to their propagation direction of graphene plasmons, the thermal rectification between the two particles is observed. The performance of the radiative thermal diode can be actively adjusted through tuning the chemical potential or changing the drift currents in the graphene sheet. With a large drift velocity and a small chemical potential, a perfect radiative thermal diode with a rectification coefficient extremely approaching to 1 can be achieved within a wide range of the interparticle distance from near to far-field. The dispersion relations of the graphene plasmons are adopted to analyze the underlying physics of the rectification effect. In addition, due to the wide band characteristic of the nonreciprocal graphene plasmons, the drift-biased graphene can act as a universal platform for the thermal rectification between particles. The particles with a larger particle resonance frequency are much more preferred to produce a better thermal diode. This technology could find broad applications in the field of thermal management at nanoscale.

## I. INTRODUCTION

A thermal diode is a thermal rectification device that preferentially facilitates heat transfer in one direction. It can be developed to be used in electronics cooling and energy conversion systems. Different studies have predicted thermal rectification between two nonlinear one-dimensional lattices [1], at the silicon-polyethylene interface [2], in asymmetric graphene nanoribbons [3], and in quantum dots [4]. These studies have considered phonon- and electron-mediated thermal transport in order to demonstrate thermal rectification. On the other hand, radiative thermal diodes have been proposed both in the near-field or far-field. The nonlinearity in these devices are basically due to the temperature dependence of the dielectric functions of the materials, such as two types of silicon carbide bulks (SiC-3C and SiC-6H) [5], a film and a bulk of doped silicon with different doping levels [6], and phase change materials [7,8] whose optical properties undergo a real bifurcation around their critical temperature [9]. Although these devices display relatively high rectification coefficients in near-field the scarcity of these phase change materials could limit the development and the operating range of this technology.

---

* Corresponding author. Tel.: +86-451-86412674; E-mail address: yihongliang@hit.edu.cn

Recently, near-field radiative heat transfer (NFRHT) between particles in the presence of a substrate has been attracting people's attention [10-15]. It is shown that when the particles are put in close proximity of the substrate the surface waves supported on the interface would provide a new channel to the energy transfer between particles, significantly enhancing the heat exchange. Due to the dominating role of the substrate in the near-field, we expect to adopt a substrate supporting non-reciprocity light propagation as a platform to rectify the radiative heat flux exchanged between two bodies. This new way would go beyond the framework of utilizing the temperature dependence of the dielectric functions of the materials to achieve the thermal rectification. In conventional photonic systems, light propagation is constrained by the Lorentz reciprocity law [16] which means that if the source and receiver are interchanged the level of received signal remains the same [17]. Recently, tremendous efforts have been made to permit nonreciprocal light propagation, enabling new applications in sub-diffractive nanophotonics, including sensing, imaging, and computing. The well-known way to realize the unidirectional propagation is by using a static magnetic field bias that creates a gyrotropic nonreciprocal response [18,19]. Most recently, the strong asymmetry in the propagation of surface waves of the magneto-optical material (InSb) has been used to rectify the radiative flux between two magneto-optical nanoparticles which are placed close to the magneto-optical substrate (InSb) [15]. A rectification coefficient up to 0.9 has been predicted. However, the unavoidable need of magnetic materials under strong bias fields significantly lessens the practical interest of this solution [20,21]. A different approach to obtain nonreciprocal surface waves is to apply drift current bias to the host surface [20-24]. Graphene has recently opened a new way in this context [20,21] thanks to its ultra-high electron mobility and large drift velocities close to the Fermi velocity ( $v_0 = 10^8$ cm/s ) [25,26]. It has been theoretically shown that drift-biased graphene can enable a broadband subwavelength "one-way" surface plasmons polaritons (SPPs), and especially remarkably the drift-current biasing can change the graphene conductivity dispersion and boost the propagation length of the graphene plasmons [20,21]. This may open new venues to collimate, steer, and process plasmons immune to backscattering. It is also worth noting that a drift current bias is fundamentally different from the standard electrical tuning of the graphene response through the control of the Fermi level [27,28].

In this work, aiming to explore a new platform to realize a radiative thermal diode we adopt the drift-biased graphene to rectify the radiative flux between nanoparticles. In Section II, we introduce the geometry of our system, and give the expression of the heat flux between the two nanoparticles in the presence of a drift-biased graphene sheet. Section III is dedicated to show the performance of the radiative thermal diode. In Section IV, the nonreciprocal dispersion relations of the graphene plasmons are analyzed to explain the underlying physics of the rectification effects. Section V examines the influence of the particle position on the rectification effects. In Section VI, we elucidate that the drift-biased graphene sheet can relax restrictions on the materials of the particles. Finally, in Sec. VII, we give some concluding remarks on this work.

## II. THEORETICAL ASPECTS

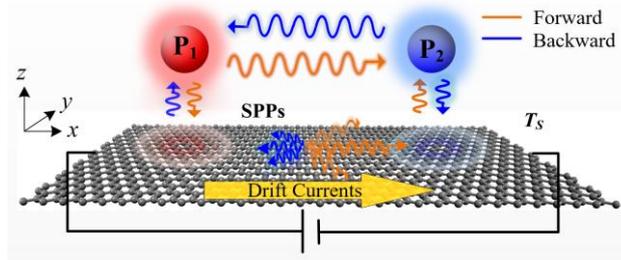

FIG. 1. Schematic of the radiative thermal diode. Two nanoparticles placed above a suspended drift-biased graphene sheet are the two terminals for the thermal diode. A static voltage generator induces drift currents along the $x$ axis in a graphene sheet. The two particles have temperatures $T_1$ and $T_2$ and the background has temperature $T_S$. With $T_1 > T_2 = T_S$ we get the forward heat flux $\Phi_F$ from $P_1$ to $P_2$. On the contrary, with $T_1 = T_S < T_2$ the backward heat flux $\Phi_B$ from $P_2$ to $P_1$ is obtained.

To start, let us consider the radiative heat transfer (RHT) between particle 1 ($P_1$) and particle 2 ($P_2$) in the presence of a suspended graphene sheet as shown in Figure 1. We assume that the propagation of the SPPs supported by the graphene sheet is driven by a direct electric current supplied by a voltage generator. By using the self-consistent field approach in the presence of drifting electrons with velocity $\mathbf{v}=v_d\mathbf{x}$ the graphene conductivity associated with a longitudinal excitation (with in-plane electric field directed along $x$) is written as the following form [20,24],

$$\sigma^d(\omega, k_x) = (\omega/\tilde{\omega})\sigma(\tilde{\omega}) \quad (1)$$

where $\tilde{\omega}=\omega-k_x v_d$ is the Doppler-shifted frequency and $k_x$ is the wavenumber along the $x$ direction. $\sigma$ is the graphene conductivity in the absence of drift currents or magnetic bias, which is modeled by the Bhatnagar-Gross-Krook (BGK) approach derived in [29] that rigorously takes the intrinsic nonlocal response of graphene into account. Intuitively, reversing the propagation direction equates to changing the sign $k_x$ and therefore the sign of the Doppler shift, i.e., $\sigma^d(\omega, k_x) \neq \sigma^d(\omega, -k_x)$ for $v_d \neq 0$, hence a nonreciprocal SPPs on the graphene sheet. We stress that the drift currents indeed generate a nonreciprocal SPPs on the graphene sheet, hence an asymmetry in the power transferred along $x$ direction between the nanoparticles. The relaxation time $\tau$ in the conductivity model is chosen as $10^{-13}$ s (following Ref. [30],). Note that drift velocities close to the Fermi velocity ($v_F \approx 10^8$ cm/s ) have been experimentally reported in graphene samples suspended in free space [31] which is selected as the maximum drifting velocity. We thus have the relation $v_d = f\, v_F$ with $f$ being the velocity ratio ranging from 0 to 1.

The two nanoparticles with a radius of $R = 5$ nm are the two terminals of the radiative thermal diode. We assume that they are isotropic, linear, and nonmagnetic so that the dipole approximation is valid [10-15]. Note also that in this scenario the radiation correction is negligible and the single-scattering approximation is valid. The two particles have temperatures $T_1$ and $T_2$ and are placed at positions of $\mathbf{r}_1 = (x_1, y_1, z)^T$ and $\mathbf{r}_2 = (x_2, y_2, z)^T$, respectively, as depicted in Fig. 1. The separation between the particle and the graphene sheet, $z$, can be controlled with a moving nanotip or with a dielectric spacer [32-34]. The background consisting of the photon gas above the graphene and the graphene itself are assumed to have the temperature $T_S$. According to the framework of fluctuational electrodynamics (FE), the heat flux $\Phi_i$ transferred from particle $j$ to particle $i$ can be expressed in terms of the Green's function (GF) [10-15,35],

$$\Phi_i = 4\int_0^{+\infty}\frac{d\omega}{2\pi}k_0^4\left[\mathrm{Im}(\alpha)\right]^2\left[\Theta(T_j)-\Theta(T_i)\right]\mathrm{Tr}\left[\mathbb{G}_{ij}\mathbb{G}_{ij}^*\right], \qquad (2)$$

where $k_0 = \omega/c$ is the free-space wavevector. $\alpha$ denotes the particle's electric frequency-dependent polarizability which can be written in the well-known Clausius-Mossoti form of $\alpha(\omega)=4\pi R^3(\varepsilon-1)/(\varepsilon+2)$ (quasi-static limit) [36] with $\varepsilon$ being the electric permittivity of the particle. $\Theta(T) = \hbar\omega/(\exp(\hbar\omega/k_B T)-1)$ denotes the thermal part of the mean energy of a harmonic oscillator with respect to a temperature of $T$. In this work, the quantities we concern are heat fluxes in two opposite directions as shown in Fig. 1: (i) "forward" heat flux $\Phi_F$ from P$_1$ to P$_2$ with $T_1 > T_2 = T_S$, and (ii) "backward" heat flux $\Phi_B$ from P$_2$ to P$_1$ with $T_1 = T_S < T_2$. To quantify the rectification in this system we introduce the rectification coefficient $\eta = |\Phi_F - \Phi_B|/\max([\Phi_F, \Phi_B])$.

$\mathbb{G}$ is the GF of this system which is the sum of the vacuum GF $\mathbf{G}^0$ and the scattering part $\mathbf{G}^{sc}$ accounting the contributions from the suspended graphene sheet, viz., $\mathbb{G} = \mathbf{G}^0 + \mathbf{G}^{sc}$. Of course, the vacuum GF for the two isolate particles is reciprocal and depends only on the interparticle distance $d$ ($|\mathbf{r}_1 - \mathbf{r}_2|$) which is given by,

$$\mathbf{G}^0 = \frac{e^{ik_0 d}}{4\pi k_0^2 d^3}\begin{pmatrix} a & 0 & 0 \\ 0 & b & 0 \\ 0 & 0 & b \end{pmatrix}, \qquad (3)$$

where $a = 2 - 2ik_0 d$ and $b = k_0^2 d^2 + ik_0 d - 1$. We can expect that any asymmetry in the heat flux,

viz., the rectification, exists only when $\mathbf{G}_{12}^{sc} \neq \mathbf{G}_{21}^{sc}$. Due to the nonreciprocal feature of the SPPs. the drift-biased graphene sheet can be regarded as a 2D anisotropic surface whose scattering GF is expressed as [37,38],

$$\mathbf{G}^{sc}(\mathbf{r}_i,\mathbf{r}_j) = \frac{i}{8\pi^2} \int_{-\infty}^{\infty} dk_x \int_{-\infty}^{\infty} \left( r_{ss}\mathbf{M}_{ss} + r_{ps}\mathbf{M}_{ps} + r_{sp}\mathbf{M}_{sp} + r_{pp}\mathbf{M}_{pp} \right) e^{ik_x(x_i-x_j)} e^{ik_x(y_i-y_j)} e^{ik_z|z_i+z_j|} dk_y, \quad (4)$$

where $r$ is the reflection coefficient related to incident '$s$' and '$p$' polarized waves [37,39]. In the numerical calculations we neglect the depolarization terms $r_{sp}$ and $r_{ps}$ because they turn out to be negligibly small.

$$r_{ss} = \frac{-\eta_0 \sigma'_{yy} \left( 2Z^p + \eta_0 \sigma'_{xx} \right) + \eta_0^2 \sigma'_{xy} \sigma'_{yx}}{\left( 2Z^s + \eta_0 \sigma'_{yy} \right)\left( 2Z^p + \eta_0 \sigma'_{xx} \right) - \eta_0^2 \sigma'_{xy} \sigma'_{yx}}, \quad (5a)$$

$$r_{pp} = \frac{-\eta_0 \sigma'_{xx} \left( 2Z^s + \eta_0 \sigma'_{yy} \right) + \eta_0^2 \sigma'_{xy} \sigma'_{yx}}{\left( 2Z^s + \eta_0 \sigma'_{yy} \right)\left( 2Z^p + \eta_0 \sigma'_{xx} \right) - \eta_0^2 \sigma'_{xy} \sigma'_{yx}}, \quad (5b)$$

where $\eta_0$ is the free-space impedance. $Z^s = k_z/k_0$, and $Z^p = k_0/k_z$. The matrices along with $r_{ss}$ and $r_{pp}$ in Eq. (4) are given as [40],

$$\mathbf{M}_{ss} = \frac{1}{k_z k_\rho^2} \begin{pmatrix} k_y^2 & -k_x k_y & 0 \\ -k_x k_y & k_x^2 & 0 \\ 0 & 0 & 0 \end{pmatrix}, \quad \mathbf{M}_{pp} = \frac{k_z}{k_0^2 k_\rho^2} \begin{pmatrix} -k_x^2 & -k_x k_y & -k_x k_\rho^2/k_z \\ -k_x k_y & -k_y^2 & -k_y k_\rho^2/k_z \\ k_x k_\rho^2/k_z & k_y k_\rho^2/k_z & k_\rho^4/k_z^2 \end{pmatrix}, \quad (6)$$

where $k_\rho = \sqrt{k_x^2 + k_y^2}$ and $k_z = \sqrt{k_0^2 - k_\rho^2}$ are the lateral and vertical wavevectors, respectively. $\boldsymbol{\sigma}'$ denotes the conductivity tensor in the wavevector space [39,41],

$$\boldsymbol{\sigma}' = \begin{pmatrix} \sigma'_{xx} & \sigma'_{xy} \\ \sigma'_{yx} & \sigma'_{yy} \end{pmatrix} = \frac{1}{k_\rho^2} \begin{pmatrix} k_x^2 \sigma_{xx} + k_y^2 \sigma_{yy} + k_x k_y \left( \sigma_{xy} + \sigma_{yx} \right) & k_x^2 \sigma_{xy} - k_y^2 \sigma_{yx} + k_x k_y \left( \sigma_{yy} - \sigma_{xx} \right) \\ k_x^2 \sigma_{yx} - k_y^2 \sigma_{xy} + k_x k_y \left( \sigma_{yy} - \sigma_{xx} \right) & k_x^2 \sigma_{yy} + k_y^2 \sigma_{xx} - k_x k_y \left( \sigma_{xy} + \sigma_{yx} \right) \end{pmatrix}, \quad (7)$$

where $\boldsymbol{\sigma}$ is the conductivity tensor of the drift-biased graphene sheet modeled by the Bhatnagar-Gross-Krook (BGK) approach [29],

$$\boldsymbol{\sigma} = \begin{bmatrix} \sigma_{xx} & \sigma_{xy} \\ \sigma_{yx} & \sigma_{yy} \end{bmatrix} \quad (8)$$

Note that $\sigma_{xx} = \sigma^d(\omega, \pm k_x)$ and $\sigma_{yy} = \sigma^d(\omega, \pm k_y)$. In this work, due to the asymmetric in $\sigma_{xx}$ induced by the drift currents along the $x$ axis, it is easy to find that $r_{ss}(k_x) \neq r_{ss}(-k_x)$ and $r_{pp}(k_x) \neq r_{pp}(-k_x)$, hence $\mathbf{G}_{12}^{sc} \neq \mathbf{G}_{21}^{sc}$. We can thus base on this asymmetric feature to achieve the rectification effect between the two particles.

## III. THERMAL RECTIFICATION BASED ON THE DRIFT-BIASED GRAPHENE SHEET

We now examine the performance of the radiative thermal diode mediated by the drift-biased graphene sheet with drift currents towards $+x$ (forward biased) introduced by a longitudinal voltage. In this work, the nanoparticles are made of silicon carbide (SiC), a typical polar dielectric materials, the dielectric function of which can be described by the Drude-Lorentz model [42], $\varepsilon(\omega) = \varepsilon_\infty \left(\omega_L^2 - \omega^2 - i\Gamma\omega\right)/\left(\omega_T^2 - \omega^2 - i\Gamma\omega\right)$ with high-frequency dielectric constant $\varepsilon_\infty = 6.7$, longitudinal optical frequency $\omega_L = 1.83 \times 10^{14}$ rad/s, transverse optical frequency $\omega_T = 1.49 \times 10^{14}$ rad/s, and damping $\Gamma = 8.97 \times 10^{11}$ rad/s. Note that the expression of the electric polarizability predicts a nanoparticle resonance frequency $\omega_{np}$ corresponding asymptotically to the condition $\varepsilon(\omega) + 2 = 0$, which for SiC gives $\omega_{np} = 1.756 \times 10^{14}$ rad/s.

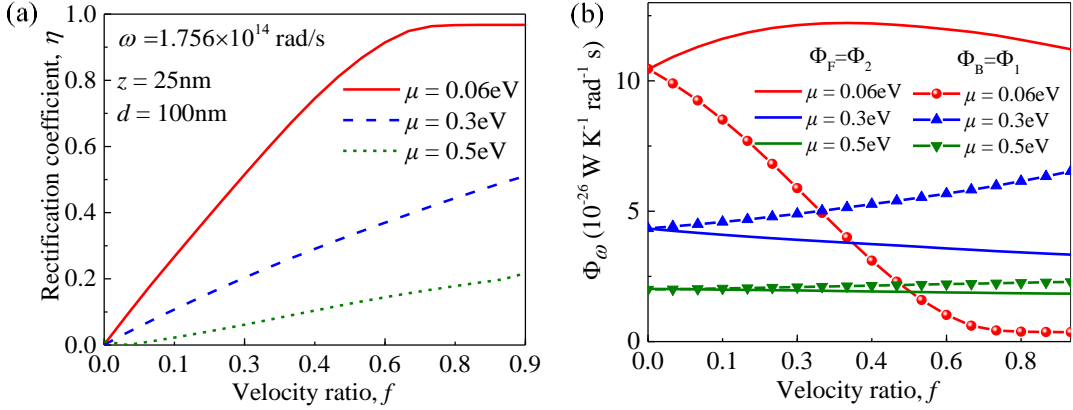

FIG. 2. (a) Rectification coefficient with respect to the velocity ratio for a chemical potential of 0.06, 0.3, and 0.5 eV, respectively. (b) Forward and backward spectral heat fluxes versus the velocity ratio at $\omega = 1.756 \times 10^{14}$ rad/s. The parameters are $\tau = 0.1$ ps, $T_G = 300$ K, $d = 100$ nm, $z = 25$ nm.

The graphene sheet coincides with the $xoy$ plane. The two particles are placed at positions of $\mathbf{r}_1 = (-50, 0, 25)^T$ nm and $\mathbf{r}_2 = (50, 0, 25)^T$ nm, respectively. The surroundings are kept at a temperature of $T_s = 300$ K. To evaluate the performance of the radiative thermal diode, two scenarios are considered, viz., forward heat flux $\Phi_F$ from $P_1$ to $P_2$ with $T_1 = 350$ K and $T_2 = 300$ K, and backward heat flux $\Phi_B$ from $P_2$ to $P_1$ with $T_2 = 350$ K and $T_1 = 300$ K, respectively. We now have the relations $\Phi_F = \Phi_2$ and $\Phi_B = \Phi_1$. The phenomenological electron relaxation time, and the temperature of the graphene sheet are set as $\tau = 0.1$ ps and $T_G = 300$ K, respectively. The above parameters are used in this work unless otherwise stated.

Figure 2(a) presents the rectification coefficient $\eta$ as a function of the velocity ratio $f$ of the drift currents for three values of chemical potentials, namely, 0.06, 0.3, and 0.5 eV, respectively. Note that $\eta$ ( $\left|\Phi_F(\omega_{np}) - \Phi_B(\omega_{np})\right|/\max([\Phi_F(\omega_{np}), \Phi_B(\omega_{np})])$ ) is based on the spectral heat flux at the SiC nanoparticle resonance frequency $\omega_{np}$. Not surprisingly, for an unbiased graphene with $f = 0$, due to the reciprocal SPPs on the graphene sheet, the rectification coefficient equals to 0.

While for $f > 0$, due to the fact that drift currents induce the nonreciprocal SPPs the thermal rectification effect is produced. We further observe that with an increase in $f$, $\eta$ increases at these three chemical potentials, indicating an increasing rectification effect. In particular, for $\mu = 0.06$ eV, when $f$ is bigger than 0.7 $\eta$ can be as high as 0.967 which is extremely close to 1.0, indicating that a nearly-perfect radiative thermal diode is achieved. While for $\mu = 0.3$ and 0.5 eV, the rectification effects are lower than that for $\mu = 0.06$ eV. Figure 2(b) gives the net spectral heat flux received by P$_1$ and P$_2$ corresponding to the backward heat flux $\Phi_B$ and forward heat flux $\Phi_F$, respectively. We observe intuitively the asymmetry of these two curves. For a chemical potential of 0.06 eV, we have $\Phi_F > \Phi_B$. With an increase in $f$, $\Phi_F$ remains at a high value, while the reverse heat flux $\Phi_B$ drops significantly. Finally, as $f > 0.7$, a biggest contrast between $\Phi_F$ and $\Phi_B$ is obtained, which confirms the nearly-perfect thermal rectification effect as shown in Fig. 2(a). For a big $\mu$, different from the case for a small $\mu$, the relation between $\Phi_F$ and $\Phi_B$ is reversed to $\Phi_B > \Phi_F$, which means that the relation between forward and backward heat fluxes depends on the value of the chemical potential. This phenomenon is discussed hereafter. We further see that for $\mu = 0.3$, and 0.5 eV, the backward flux increases with respect to $f$, while the forward flux decreases. Referring to Eq. (2) and the formulation of $\eta$, we note that the temperature difference between the two terminals would change the net heat flux between the two terminals, but make no influence on the rectification coefficient of the proposed thermal diode.

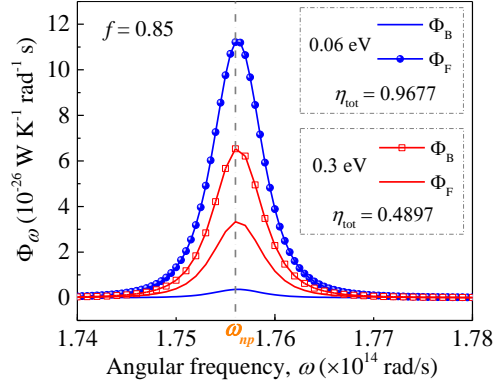

FIG. 3. Spectral distributions of the forward and backward heat fluxes for chemical potentials of 0.06 eV and 0.3 eV at $f = 0.85$.

Figure 3 presents the spectral distributions of the forward and backward heat fluxes for chemical potentials of 0.06 eV and 0.3 eV at $f = 0.85$. It is shown that the spectral heat flux at $\omega_{np}$ dominates the RHT. In addition, we can always observe the asymmetric RHT between the two particles within the spectral band under consideration. This is due to the fact that the dispersion relations of the graphene SPPs covers a wide spectral band. We further see that the rectification coefficients defined by the total heat flux ($\eta_{tot} = \left| \int_\omega \Phi_F - \Phi_B \right| / \max([\int_\omega \Phi_F, \int_\omega \Phi_B])$) for 0.06 eV and 0.3 eV at $f = 0.85$ equal to 0.9677 and 0.4897, respectively, which are the same as those predicted by the spectral heat flux at $\omega_{np}$ as shown in Fig. 2(a). This means that the rectification behavior at $\omega_{np}$ is equivalent to that for the total heat flux. Hence, in this work all the analysis and discussions are based on the spectral heat flux $\Phi_\omega$ at $\omega_{np}$.

**IV. UNDERLYING PHYSICAS OF THE RECTIFICATION**

Eqs. (2) and (4) show that effects of graphene sheet on the heat transfer between particles are reflected in the reflection coefficient matrix and the scattering GF. We thus employed these two physical quantities to explore the underlying physical mechanisms for this nonreciprocal thermal radiation phenomenon.

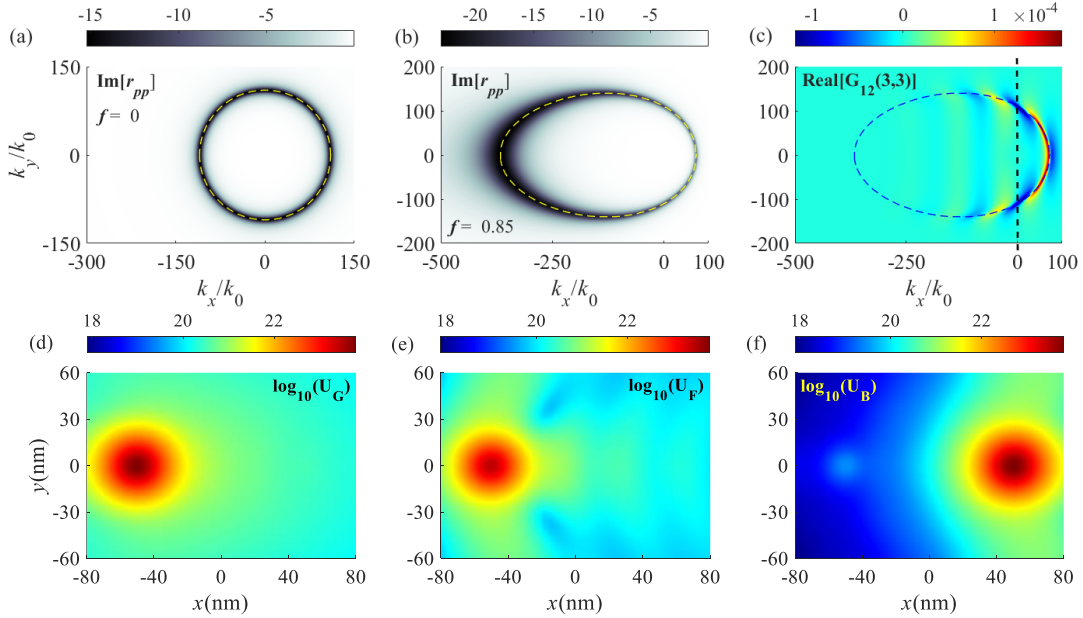

FIG. 4. Wavevector contours of the imaginary part of the complex reflection coefficient Im($r_{pp}$) for a velocity ratio of (a) $f = 0$, and (b) $f = 0.85$. (c) Wavevector contours of the real part of the reflected Green's function. Spatial contours of the electric field energy density U at $z = 12.5$nm for (d) the graphene sheet without drift current $\log_{10}(U_G)$, and with a drift current for (e) the forward case $\log_{10}(U_F)$ and (f) the backward case $\log_{10}(U_B)$. In panels (a)~(c), the dashed lines denote the dispersion relations of the SPPs supported on the drift-biased suspended graphene sheet. The blacked dashed line in panel (c) separate the left and right half-space. For panels (d)~(f), the higher and lower temperatures are kept at 300 K and 0.5 K, respectively. A velocity ratio of 0.85 is used for the drift-biased graphene sheet. The other parameters are $d = 100$ nm, $z = 25$ nm, $\omega = 1.756 \times 10^{14}$ rad/s, and $\mu = 0.06$ eV.

For a chemical potential of 0.06 eV, Figures 4(a) and 4(b) show the distributions of the reflection coefficient $r_{pp}$ of the graphene sheet with $f = 0$ and 0.85, respectively. The yellow curve shows the dispersion relation which is given by [38] $2\kappa_0^2\eta_0\left(\sigma''_{xx} + \sigma''_{yy}\right) - 2\eta_0\left(k_x^2\sigma''_{xx} + k_y^2\sigma''_{yy}\right) + \kappa_0 k_z\left(4 + \eta_0^2\sigma''_{xx}\sigma''_{yy}\right) - \eta_0^2\sigma''_{xx}\sigma''_{yy}k_0 k_z = 0$, where the superscript $''$ signifies the imaginary components of the conductivity, and $\eta_0$ is the free-space impedance. This equation can be numerically solved with conventional root-finding algorithms in the complex plane [39]. We can see that the dispersion curves match well with the resonance region in the $r_{pp}$ contour. It is shown in Fig. 4(a) that, since there is not drift current in the graphene sheet for $f = 0$, the dispersion curve of the SPPs is a usual isotropic circle. While with $f > 0$, the drift bias breaks the symmetry of SPPs with positive and negative $k_x$, leading to an effectively anisotropic two-dimensional medium. In particular, for a large drift current with $f = 0.85$, the reflection coefficient distributes an eggs-like shape as shown in Fig. 4(b). It is because that, when large drift currents propagate towards $+x$, the dispersion in the negative $k$ half-space (propagating towards $-x$) is significantly dragged to large wavevectors, indicating high $k$ states against the drift. We stress that it is indeed the asymmetry of SPPs that results in the nonreciprocal RHT, hence a rectification effect.

However, the behavior of the reflection coefficient only tells us the characteristic of an isolate graphene sheet, but cannot show how it interacts with the two particles. We now turn our attention to the GF as shown in Figure 4(c), which is greatly distinguished from the $r_{pp}$ plot. Note that non-zero values in Re[$G_{12}$] indicate the efficient interaction of the surface waves at the two particle positions. We see that GF show much more non-zero values in the positive $k_x$ half-space than the negative one, especially along the line of $k_y = 0$. This implies that most surface waves in the $+k_x$ half-space contribute to the forward energy transfer between particles. However, those plasmons in negative $k_x$ half-space barely make contributions to the backward energy transfer. Mathematically, the term $e^{ik_z|z_i+z_j|}$ in the formulation of GF accounts for the evanescent feature of the SPPs on the graphene sheet along the vertical direction. Based on this fact, we thus expect that not all surface states participate in the scattering enhancement process of the RHT between particles. For a drift-biased graphene sheet with $f = 0.85$, although plenty of surface waves show high $k$ features in the $-k_x$ region, they are significantly more confined and lossy than their $+k_x$ counterparts. At a location of 25 nm above the graphene sheet, we see in Fig. 4(c) that all the right surface states on the dispersion curves can efficiently couple to the waves emitted by the particles, but most of the left high $k$ components are filtered by free space and cannot efficiently couple to the waves emitted by the particles. Consequently, the backward heat flux $\Phi_B$ from $P_2$ to $P_1$ (towards $-x$) is suppressed to a small value. While the forward heat flow $\Phi_F$ from $P_1$ to $P_2$ (towards $+x$) is significantly larger than the backward one. Finally, under such parameter settings of $\mu = 0.06$ eV, $z = 25$ nm, $d = 100$ nm, $R = 5$ nm, and $f = 0.85$, a very high rectification coefficient 0.967 is achieved, indicating a nearly-perfect radiative thermal diode.

In order to illustrate the physical process more intuitively, we display the spatial distributions of the radiated electric field energy density $U_e(\mathbf{r},\omega) = \frac{2\varepsilon_0^2}{\pi\omega}\sum_j \chi_j \Theta(\omega,T_j) \mathrm{Tr}\left[\mathbb{Q}_{rj}\mathbb{Q}_{rj}^*\right]$ where $\mathbb{Q}_{rj} = \omega^2\mu_0\left(G_0^{rj} + G_R^{rj}\right)\mathcal{G}$ [28], in the mid-plane between particle and graphene sheet at $z = 12.5$ nm for $f = 0$ in Figure 4(d), and for $f = 0.85$ in Figures 4(e) for forward case and 4(f) for backward case, respectively. The energy distribution for the case with a non-drift-biased graphene sheet is homogeneous and isotropic as shown in Fig. 4(d). While we see that the drift currents in the graphene sheet significantly modifies the energy distributions in the three-body system, exhibiting distinctive inhomogeneity. For the forward case, energy flows smoothly towards P2 as observed in Fig. 4(e) which can be viewed as the 'on' mode. On the contrary, we observe in Fig. 4(f) that the energy emitted by the $P_2$ barely reaches the $P_1$, which can thus be regarded as the 'off' mode. We stress that although the SPPs of the drift-biased graphene exhibits nonreciprocal features as shown in Fig. 4(b), the dispersion relations still show a closed shape, which means that SPPs are still supported in every direction. However, due to the evanescent feature of the surface waves, most part of the confined and lossy SPPs in $-x$ direction can be filtered by the free space. The supported SPPs are therefore immune to the backward heat transfer between particles and thus may make an ideal platform for the radiative thermal diode.

In order to further understand the trends of the curves for different $\mu$ in Fig. 2, we plot the dispersion relations of the graphene sheet for different $f$ at $\omega_{np}$ in Figures 5(a) and 5(b). Note that the dispersion relation is also the isofrequency contour (IFC) of the SPPs. In the absence of any biasing at $f = 0$, since all supported SPPs exhibit identical characteristics independently of their propagation direction, SPPs is isotropic and thus possesses a circular IFC. With a smaller chemical potential, the graphene sheet supports higher $k$ surface states. We thus observe a larger heat flux for a smaller $\mu$ in Fig. 2(b) at $f = 0$. As the voltage is applied, the IFC is dragged towards $-k_x$ half-space, which is opposite to the direction of the drift currents ($+x$). Increasing the drift currents further boosts the asymmetry of the IFC. The negative $k_x$ half-space is dragged much more obvious than the positive $k_x$ half-space. In Figures 5(c) and 5(d), the two intersections between the line of $k_y = 0$ and the IFC, viz., $k_{x,R}$ and $k_{x,L}$, are used to characterize the asymmetry degree of the IFC. For a small value $\mu$ at 0.06 eV, great deviation between $k_{x,R}$ and $k_{x,L}$ is observed which is ranged from 0 to -300 as $f$ increases from 0 to 0.85. With an increase in $f$, due to the low $k$ feature the surface states in the positive $k_x$ half-space can always couple to the waves emitted by the particles. We thus see that the forward heat flux (towards $+x$) maintains at a large value as shown in Fig. 2(b). While with rapid increases in $|k_{x,L}|$ the surface states in the negative $k_x$ half-space become more and more confined and lossy, and thus cannot participate in the RHT between particles any more. Consequently, the backward heat flux (towards $-x$) decrease rapidly as observe in Fig. 2(b). Finally, we observe the large rectification effect as shown in Fig. 2(a). While for a larger $\mu$ of 0.3 eV, due to the much smaller wavevectors and the low-lossy feature, all the surface states can contribute to the RHT between particles for any values of $f$. In other words, no surfaces states are filtered by the free space. The decrease in $k_{x,R}$ results in the decline trend of the forward heat flux ( the blue line in Fig. 2(b)). While the increase in $k_{x,L}$ explains the climb trend of the backward heat flux (the dashed blue line in Fig. 2(b)). These two opposite trends lead to the rectification effect. However, as shown in Fig. 5(d) for a large $\mu$ the deviation between $k_{x,R}$ and $k_{x,L}$ is small (ranging from 0 to -9), hence a small difference between the backward and forward heat flux is observed in Fig. 2(b).

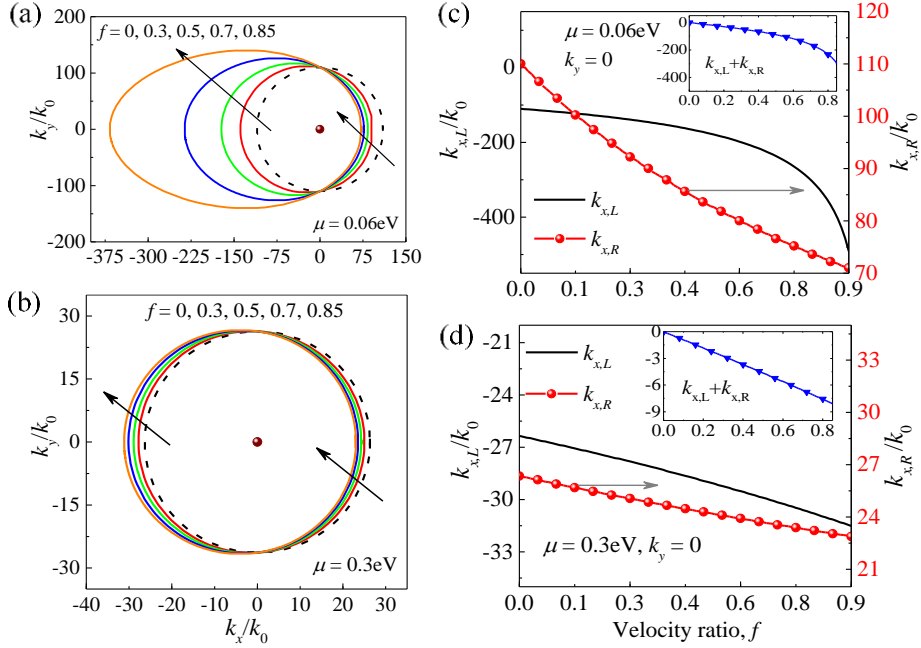

FIG. 5. Dispersion relations of the SPPs supported by the graphene sheet for different $f$ at $\omega_{np}$ for (a) $\mu = 0.06$ eV, and (b) $\mu = 0.3$ eV. Wavevector of the SPPs towards $+x(k_{x,R})$ and $-x$ ($k_{x,L}$) at the line of $k_y = 0$ for (c) $\mu = 0.06$ eV, and (d) $\mu = 0.3$ eV.

With $f$ ranging from 0 to 0.9 and $\mu$ varied from 0 to 0.5 eV, and meanwhile the other parameters kept the same as those in Fig. 2, forward and backward heat fluxes are presented in Figures 6(a) and 6(b), respectively. Great distinctions are observed between $\Phi_F$ and $\Phi_B$. With an increase in $\mu$ at a fixed $f$, both the forward and backward heat fluxes increase firstly and then decrease. The rectification coefficients defined by $\eta = (\Phi_F - \Phi_B)/\max([\Phi_F, \Phi_B])$ is plotted in Figure 6(c). It is shown that the sign of the $\eta$ could be either positive or negative. The black dashed line in Fig. 6(c) marks the region for $\eta = 0$. In the left or red region where the chemical potential is small, the forward heat flux is significantly higher than the backward one. While in the right or blue region, the backward heat flux is higher than the forward one. The common feature is that in the case of $d = 100$ nm and $z = 25$ nm with a larger drift current the rectifier coefficient is higher. In addition, we stress that the graphene sheet with a drift current cannot necessarily generate asymmetric heat flow. For instance, under the parameter condition where the black line is located, the forward heat flux equals to the backward one, indicating that the rectification effect is not activated. This is because that, under such condition the surface waves in $+k_x$ or $-k_x$ regions contribute equally to the thermal radiation between the two particles.

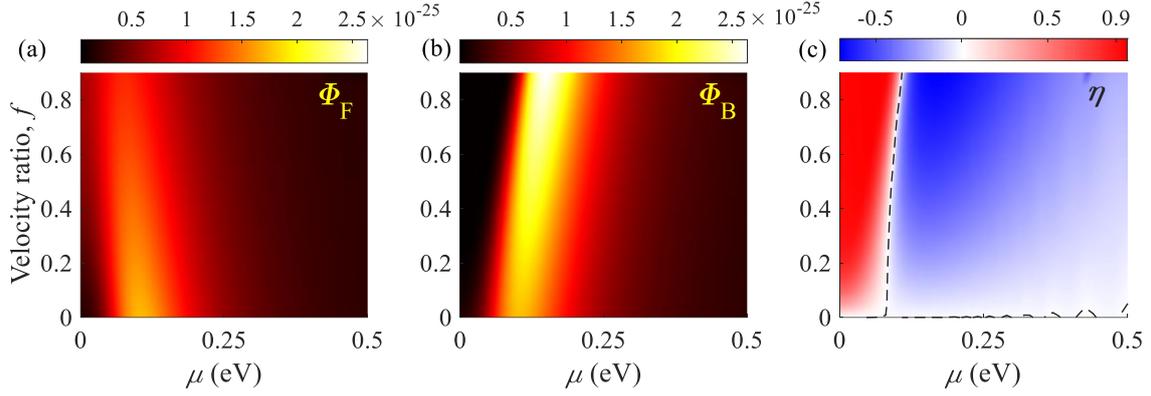

FIG. 6. (a) Forward heat flux $\Phi_F$, (b) backward heat flux $\Phi_B$, and (c) rectification coefficients defined by $\eta = (\Phi_F - \Phi_B)/\max([\Phi_F, \Phi_B])$ with respect to different drift velocities and chemical potentials. The black dashed line marks the zero value in the rectification coefficient.

## V. RADIATIVE THERMAL DIODE AT DIFFERENT PARTICLE POSITIONS

So far, the performance and the corresponding underlying physics of the radiative thermal diode when the line connecting the two particles is parallel to the $k_x$ axis ($k_y = 0$) have been examined. The eggs-like IFC in Fig. 5(a) implies that the nonreciprocal characteristics of the SPPs is significantly varied with the direction. Meanwhile, the propagation length of the plamons would also play a crucial role on the heat transfer between particles. One can thus expect that the positions of $P_1$ and $P_2$ could change the performance of the radiative thermal diode. The particles can be placed in two ways: (i) keeping the $d$ unchanged, while changing the relative positions between particles and graphene sheet; (ii) varying the interparticle distance $d$.

Now, let us firstly consider the first scenario. By keeping an interparticle distance of $d = 100$ nm, we simultaneously rotate the two particles counterclockwise by an angle of $\theta$ as depicted in the inset of the Figure 7(a). The other parameters are set as $z = 25$ nm, $\omega = 1.756 \times 10^{14}$ rad/s, $\mu = 0.06$ eV, and $f = 0.85$. We thus have two particles placed at positions of $\mathbf{r}_1 = [-d\cos(\theta)/2, -d\sin(\theta)/2, z]^T$ and $\mathbf{r}_2 = [d\cos(\theta)/2, d\sin(\theta)/2, z]^T$, respectively. Results of the rectification coefficient, and the corresponding forward and backward heat fluxes are presented in Fig. 7(a). We see that $\eta$ is almost unchanged when $\theta$ is varied from 0° to 75°. By further increasing the rotation angle, $\eta$ decreases rapidly and finally equals to zero at 90°, indicating the vanish of the rectification effect. The trend of the rectification coefficient curve results from the behaviors of the forward and backward heat transfer. A non-monotonic trend with respect to $\theta$ is observed in the curve for the forward heat flux. While the curve for the backward heat flux rises monotonically. The underlying physics of this phenomenon can be understood by analyzing the behavior of the SPPs between the two particles. Figure 7(b) qualitatively shows how rotation affects the SPPs in the two half-spaces. The closed yellow line represents the IFC of the SPPs at $f = 0.85$. The segments of the IFC covered by the blue and green cuboid represent the forward and backward

SPPs, viz., SPPs towards P$_2$ and P$_1$, respectively. The arrows in Fig. 7(b) perpendicular to the IFC qualitatively illustrate the propagation directions of the graphene plasmons [21].

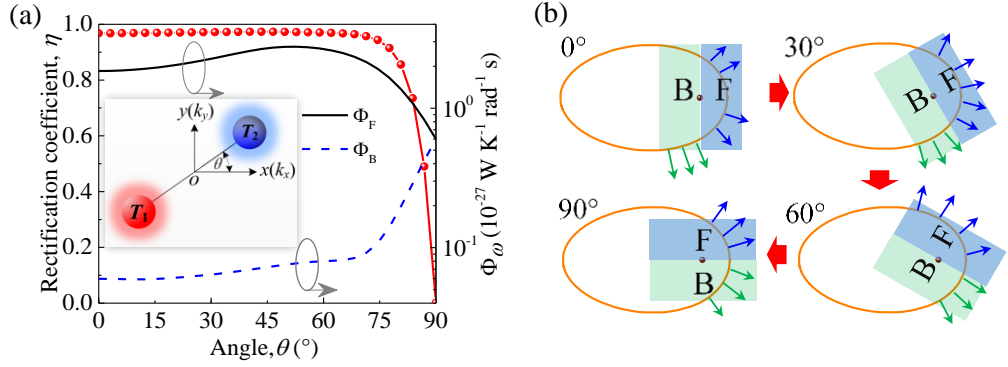

FIG. 7. Radiative thermal diode at different particle positions with a fixed interparticle distance of $d = 100$ nm. (a) Rectification coefficients, forward and backward radiative heat fluxes with respect to the rotation angle $\theta$. The inset shows the schematic of the rotation. (b) Forward and backward SPPs at different $\theta$. The yellow line represents the IFC. The segments of the IFC covered by the blue and green cuboids represent excitation of the forward and backward SPPs, respectively. The blue and green arrows perpendicular to the IFC qualitatively illustrate the directions of the waves. The black dashed line divides the regions into the top right one and the left bottom one, representing the forward and backward modes, respectively. For these two panels, the other parameters are set as $z = 25$ nm, $\omega = 1.756 \times 10^{14}$ rad/s, $\mu = 0.06$ eV, and $f = 0.85$.

We see that, with an increase in $\theta$, the SPPs towards P$_2$ (forward) transform from the right region to the top one, which results in a slight increase in the wavevectors of the forward SPPs, hence a climbing trend in the black line as shown in Fig. 7(a). However, by further increasing $\theta$ to be a value larger than 60°, due to the increase in the wavevector, those waves directly propagating towards P$_2$ become much more lossy, and are thus filtered in the free space, resulting in a descending trend in the black curve. Similarly, the backward SPPs evolves from the left region to the bottom one. As has been mentioned above, since the left high $k$ modes show ultra-confinement and large lossy features at $\theta = 0°$, most of the high $k$ surface states are filtered by the free space, especially the surface states directly towards P$_1$, hence resulting in a very small value of the backward heat flux. As $\theta$ increases, more and more surface states couple to the waves emitted by the particles. However, as $\theta < 70°$, the reflected surface waves from the graphene sheet do not propagate directly to the P$_1$ as shown in Fig. 7(b), and thus contribute negligibly to the backward heat transfer, resulting in the negligible ascending trend in the blue dashed curve in Fig. 7(a). When $\theta$ goes beyond 70°, the backward heat flux increases significantly, indicating that more and more surface modes propagate directly towards P$_1$. Obviously, with the increasing in $\theta$, the regions for the forward and backward SPPs gradually approach to each other. Consequently, the asymmetry feature of the heat fluxes in the two opposite directions gradually fades and eventually disappears at $\theta = 90°$. In fact, since the drift currents move along the $x$ axis, we thus expect no rectification effect when the two particles are placed in the $y$ axis.

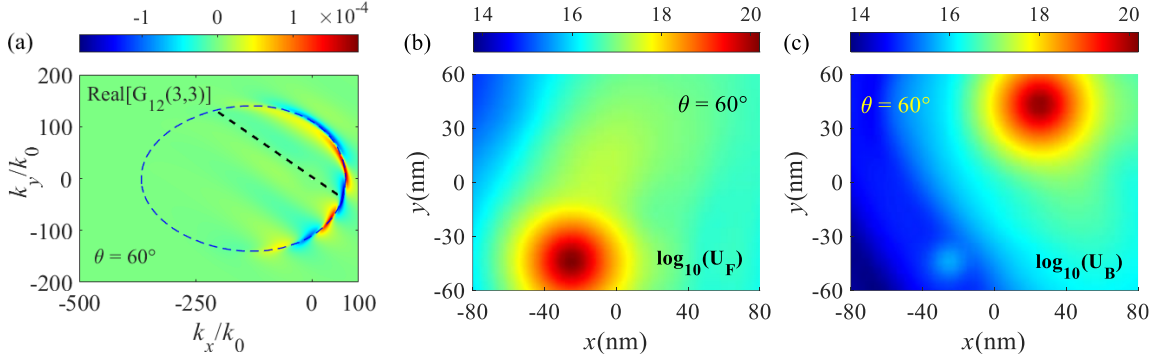

FIG. 8. (a) Wavevector contours of the real part of GF at $\theta = 60°$. The black dashed line separates the forward and backward regions. Spatial contours of the electric field energy density at $z = 12.5$ nm for the two particles rotated at $\theta = 60°$ for (b) the forward case $\log_{10}(U_F)$, and (c) the backward case $\log_{10}(U_B)$. The parameters are set as $z = 25$ nm, $\omega = 1.756 \times 10^{14}$ rad/s, $\mu = 0.06$ eV, $f = 0.85$, and $d = 100$ nm.

In order to have an intuitive understanding of the influence of rotation on the energy distribution, as a concrete example we present the GF for the two particles rotated at $\theta = 60°$ in Figure 8(a), and the spatial distributions of the radiated electric field energy density for the forward and backward cases in Figures 8(b) and 8(c), respectively. Different from the symmetrical distribution at $\theta = 0°$ as shown in Figs. 4(c), 4(e) and 4(f), GF and energy density distributions after rotation are obviously asymmetric. It can be further seen that the energy distribution between $P_1$ and $P_2$ in Fig. 8(b) is uniform and remains at a high level, indicating that the forward surface states from $P_1$ to $P_2$ participate in the forward heat transfer between particles. While for the backward case in Fig. 8(c) it is shown that the energy density around $P_1$ is obviously at a low level, which means that the energy transferred to $P_1$ from $P_2$ is restricted. This implies that the backward SPPs from P2 to $P_1$ is filtered by the free space. However, we observe big energy densities at the lower right corner. This can be qualitatively understood by the interpretation of Fig. 7(b) for $\theta = 60°$. It is shown that the green arrows point to the lower right, indicating the propagation of the surface waves towards the lower right corner.

We now turn our attention to discuss the performance of the radiative thermal diode at different interparticle distances. Figure. 9 presents the results for the rectification coefficients, forward and backward radiative heat fluxes with respect to $d$ at $\theta = 0°$. We see that $\eta$ can be further pushed up to a value extremely approximate to 1.0 when the distance is larger than 100nm, indicating a perfect radiative thermal diode. The corresponding perfect interparticle distance $d$ can range from near to far-field (0.2nm~2μm). We see that the curve of $\Phi_B$ overlaps with that of $\Phi_0$ for the isolate two particles when $d$ is below 300nm. Keep in mind that few graphene plasmons contributes to the backward heat flux. At a small distance, compared with the direct contributions to the heat flux from particle to particle, those from graphene is negligible. However, 'few waves' does not mean 'no waves'. There are indeed a little bit SPPs making interaction with particles. Due to the lower decay rate of the magnitude of the surface waves on the graphene sheet with respect to *d* than that of the waves emitted by particles, the contribution from the SPPs gets more and more prominent, hence $\Phi_B > \Phi_0$ at a large distance. For the forward heat transfer $\Phi_F$, due to the assistance with plenty of graphene SPPs, an enhancement of heat transfer with several orders of magnitude higher than that without the graphene sheet is achieved at all the considered distances. The high contrasts between $\Phi_B$ and $\Phi_F$ result in the perfect thermal diode. However, the propagation length of the forward SPPs is lower than that of the backward SPPs. This means that the amplitude of the forward SPPs decays with respect to *d* at a faster rate than that of the backward SPPs, resulting in an earlier decrease trend in $\Phi_F$ than in $\Phi_B$. Consequently, a descending trend of the curve for $\eta$ is observed when the two particles are put in the far field regime.

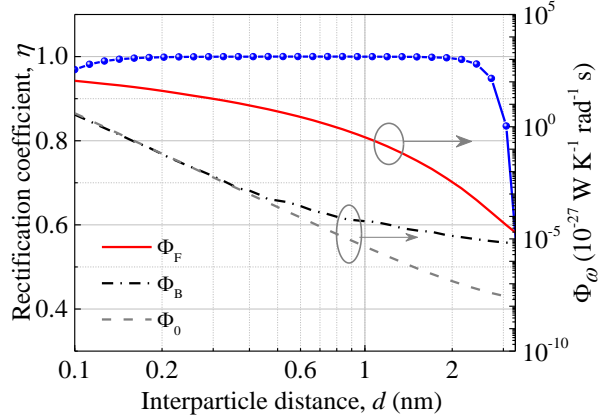

FIG. 9. Rectification coefficients, forward and backward radiative heat fluxes with respect to the interparticle distance $\theta = 0°$. The gray dashed line represents the heat flux $\Phi_0$ in the absence of the graphene sheet. The parameters are set as $z = 25$ nm, $\omega = 1.756 \times 10^{14}$ rad/s, $\mu = 0.06$ eV, and $f = 0.85$.

## VI. UNIVERSALITY OF THE DRIFT-BIASED GRAPHENE FOR THE THERMAL RECTIFICATION

Figure 10(a) shows the dispersion relations with respect to the angular frequency for the drift-biased graphene sheet with $\mu = 0.06$ eV at a velocity ratio of 0, 0.4, and 0.8, respectively. It is shown that the drift-biased graphene can support nonreciprocal SPPs within a wide spectral band, especially for $\omega > 1.0 \times 10^{14}$ rad/s. The asymmetric feature between $k_{x,L}$ and $k_{x,R}$ gets much more prominent at a larger frequency. We stress that this feature can relax the restrictions on the materials of the particles. This means that the drift-biased graphene can act as a universal platform for the thermal rectification between particles whose resonance frequency lies above $1 \times 10^{14}$ rad/s. In order to show this universality, the particles made by cBN and InP are considered. The dielectric function of cBN is described by the Drude-Lorentz model [42], $\varepsilon(\omega) = \varepsilon_\infty \left( \omega_L^2 - \omega^2 - i\Gamma\omega \right) / \left( \omega_T^2 - \omega^2 - i\Gamma\omega \right)$ with high-frequency dielectric constant $\varepsilon_\infty = 4.46$, longitudinal optical frequency $\omega_L = 2.45 \times 10^{14}$ rad/s, transverse optical frequency $\omega_T = 1.985 \times 10^{14}$ rad/s, and damping $\Gamma = 9.93 \times 10^{11}$ rad/s. While the dielectric function of InP can be described by the Drude–Lorentz model [43] in the form of $\varepsilon(\omega) = \varepsilon_\infty - \left(\omega_p^2/(\omega^2 + i\gamma_p\omega)\right) + \left(A_L\omega_L^2/(\omega_L^2 - \omega^2 - i\gamma_L\omega)\right)$ with high-frequency dielectric constant $\varepsilon_\infty = 10.01$, plasma frequency $\omega_p = 4.7 \times 10^{14}$ rad/s, lattice frequency $\omega_L = 5.73 \times 10^{13}$ rad/s, lattice vibration amplitude $A_L = 2.89$, plasma damping constant $\gamma_p = 8.85 \times 10^{13}$ rad/s, and lattice damping constant $\gamma_L = 2.09 \times 10^{12}$ rad/s. We can easily find that the particle resonance frequencies $\omega_{np}$ for cBN and InP are $2.327 \times 10^{14}$ rad/s and $1.11 \times 10^{14}$ rad/s, respectively. Results of $\eta$ with respect to $f$ at the corresponding $\omega_{np}$ for three different $\mu$ are presented in Figure 10(b). We see that with a largest $\omega_{np}$ the cBN particles produce a highest $\eta$ at any chemical potentials. While for the InP particles, due to the negligible nonreciprocal features of the graphene SPPs at $1.11 \times 10^{14}$ rad/s, the thermal rectification effect is not obvious especially for a large $\mu$. We can thus conclude that based on the drift-biased graphene platform the particles with a larger $\omega_{np}$ are much more preferred to produce a better thermal diode.

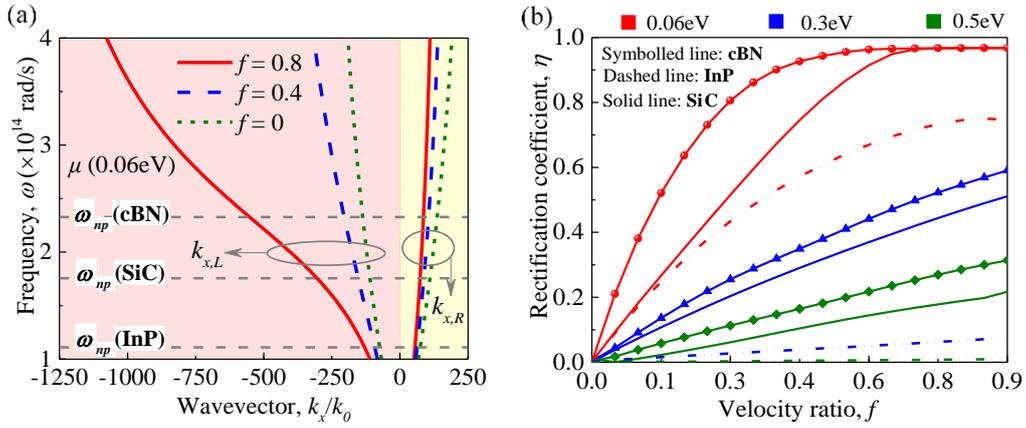

FIG. 10. (a) Dispersion relations with respect to the angular frequency for the drift-biased graphene sheet with $\mu$ = 0.06 eV at a velocity ratio of 0, 0.4, and 0.8, respectively. The regions shaded in red and yellow represented the negative and positive $k_x$ half-space, respectively. The gray dashed lines denotes the resonance frequencies for the cBN, SiC, and InP nanoparticles at $2.327 \times 10^{14}$ rad/s, $1.756 \times 10^{14}$ rad/s, and $1.11 \times 10^{14}$ rad/s, respectively. (b) Rectification coefficients between the two particles made by different materials at the corresponding particle resonance frequency. The parameters are $\tau$ = 0.1 ps, $T_G$ = 300 K, $d$ = 100 nm, and $z$ = 25 nm.

## VII. CONCLUSIONS

We have constructed a three-body system where a drift-biased graphene sheet is placed in the proximity to two particles to rectify the nanoscale radiative heat flux between two nanoparticles. In the near-field regime graphene plasmons can couple to the waves emitted by the particles, and thus modify the heat transfer between two particles. We have observed great asymmetry between the forward and backward heat flux between two particles, hence a radiative thermal diode has been achieved. This rectification mechanism is based on the asymmetry with respect to the propagation direction of nonreciprocal graphene plasmons induced by the drift currents in the graphene sheet. We have shown that the rectification coefficient can be actively tuned by changing the drift velocities and the chemical potentials. In particular, with a large drift velocity and a small chemical potential, the SPPs opposite to the direction of the drift currents are switched to higher $k$ modes, and thus filtered by the free space, indicating the off mode of the heat flux. However, the SPPs towards the drift direction change negligibly, hence an on mode of the heat flux. Thanks to the different asymmetries of the nonreciprocal graphene plasmons along different directions, the performance of the thermal diode can be varied with the positions of the particles. We have found that a perfect radiative thermal diode displaying a rectification coefficient extremely approaching to 1.0 can be achieved within a wide range of the interparticle distance from near to far-field. Finally, based on the wide band characteristic of the nonreciprocal graphene plasmons, we have shown that the drift-biased graphene can act as a universal platform for the thermal rectification between particles whose resonance frequency lies above $1\times10^{14}$ rad/s. The particles with larger particle resonance frequencies are much more preferred to produce a better thermal diode. The technology proposed in this work could find broad applications in the field of thermal management at nanoscale.

## ACKNOWLEDGMENTS

This work was supported by the National Natural Science Foundation of China (Grant No. 51706053), as well as the Fundamental Research Funds for the Central Universities (Grant No. HIT. NSRIF. 201842), and by the China Postdoctoral Science Foundation (Grant No. 2017M610208).

## REFERENCES

[1] B. W. Li, L. Wang, and G. Casati, Phys. Rev. Lett. **93**, 184301 (2004).

[2] M. Hu, P. Keblinski, and B. Li, Appl. Phys. Lett. **92**, 211908 (2008).

[3] J. N. Hu, X. L. Ruan, and Y. P. Chen, Nano Lett. **9**, 2730 (2009).

[4] D. M. T. Kuo and Y. C. Chang, Phys. Rev. B **81**, 205321 (2010).

[5] C. R. Otey, W. T. Lau, and S. H. Fan, Phys. Rev. Lett. **104**, 154301 (2010).

[6] S. Basu and M. Francoeur, Appl. Phys. Lett. **98**, 113106 (2011).

[7] Y. Yang, S. Basu, and L. P. Wang, Appl. Phys Lett **103**, 163101 (2013).

[8] K. Ito, K. Nishikawa, H. Iizuka, and H. Toshiyoshi, Appl. Phys. Lett. **105**, 253503 (2014).

[9] M. M. Qazilbash, M. Brehm, B. G. Chae, P.-C. Ho, G. O. Andreev, B.J. Kim, S.J. Yun, A.V. Balatsky, M.B. Maple, F. Keilmann, H.T. Kim, and D.N. Basov, Science **318**, 1750 (2007).

[10] J. Dong, J. Zhao, and L. Liu, Phys. Rev. B **97**, 075422 (2018).

[11] R. Messina, S.-A. Biehs, and P. Ben-Abdallah, Phys. Rev. B **97**, 165437 (2018).

[12] K. Asheichyk and M. Kruger, Phys. Rev. B **98**, 195401 (2018).

[13] Y. Zhang, M. Antezza, H. L. Yi, and H. P. Tan, Phys. Rev. B **100**, 085426 (2019).

[14] Y. Zhang, H. L. Yi, H. P. Tan, and M. Antezza, Phys. Rev. B **100**, 134305 (2019).

[15] A. Ott, R. Messina, P. Ben-Abdallah, and S.-A. Biehs, Appl. Phys. Lett. **114**, 163105 (2019).

[16] H. B. G. Casimir, Rev. Mod. Phys. **17**, 343-350 (1945).

[17] L. Deak and T. Fulop, Ann. Phys-New York **327**, 1050 (2012).

[18] L. Lu, J. D. Joannopoulos, and M. Soljacic, Nat. Phys. **12**, 626 (2016).

[19] Y. Poo, R. X. Wu, Z. F. Lin, Y. Yang, and C. T. Chan, Phys. Rev. Lett. **106**, 093903 (2011).

[20] T. A. Morgado, and M. G. Silveirinha. ACS Photon. **5**, 4253-4258 (2018).

[21] D. Correas-Serrano and J. S. Gomez-Diaz. Phys. Rev. B **100**, 081410(R) (2019).

[22] A. R. Davoyan and N. Engheta, Nat. Commun. **5**, 5250 (2014).

[23] T. Wenger, G. Viola, J. Kinaret, M. Fogelstrom, and P. Tassin, Phys. Rev. B **97**, 085419 (2018).

[24] T. A. Morgado and M. G. Silveirinha, Phys. Rev. Lett. **119**, 133901 (2017).

[25] H. Ramamoorthy, R. Somphonsane, J. Radice, G. He, C. P. Kwan, and J. P. Bird, Nano Lett. **16**, 399 (2016).

[26] M. K. Yamoah, W. M. Yang, E. Pop, and D. Goldhaber-Gordon, ACS Nano **11**, 9914 (2017).

[27] H. L. Qian *et al.*, ACS Nano **8**, 2584 (2014).

[28] Y. Liu, J. S. Zhang, H. P. Liu, S. Wang, and L. M. Peng, Sci. Adv. **3**, e1701456 (2017).

[29] G. Lovat, G. Hanson, R. Araneo, and P. Burghignoli, Phys. Rev. B **87**, 115429 (2013).

[30] M. Jablan, H. Buljan, and M. Soljačić, Phys. Rev. B **80**, 245435 (2009).

[31] V. E. Dorgan, A. Behnam, H. J. Conley, K. I. Bolotin, and E. Pop, Nano Lett. **13**, 4581 (2013)

[32] A. Y. Nikitin, P. Alonso-González, S. Vélez, S. Mastel, A. Centeno, A. Pesquera, A. Zurutuza, F. Casanova, L. E. Hueso, F. H. L. Koppens, and R. Hillenbrand, Nat. Photon. **10**, 239 (2016).

[33] J. Chen, M. Badioli, P. Alonso-González, S. Thongrattanasiri, F. Huth, J. Osmond, M. Spasenovic, A. Centeno, A. Pesquera, ´P. Godignon, A. Z. Elorza, N. Camara, F. J. García de Abajo, R. Hillenbrand, and F. H. L. Koppens, Nature (London) **487**, 77 (2012).

[34] A. Woessner, M. B. Lundeberg, Y. Gao, A. Principi, P. AlonsoGonzález, M. Carrega, K. Watanabe, T. Taniguchi, G. Vignale, M. Polini, J. Hone, R. Hillenbrand, and F. H. L. Koppens, Nat. Mater. **14**, 421 (2014).

[35] P. Ben-Abdallah, S.-A. Biehs, and K. Joulain, Phys. Rev. Lett. **107**, 114301 (2011).

[36] A. Lakhtakia, V. K. Varadan, and V. V. Varadan, Int. J. Infrared Milli. **12**, 1253-1264 (1991).

[37] A. Lakhatakia, Int. J. Infrared Milli. **13**, 161 (1992).

[38] G. W. Hanson, IEEE T. Antenn. Propag. **56**, 747 (2008).

[39] J. S. Gomez-Diaz, M. Tymchenko, and A. Alù, Opt. Mater. Express **5**, 2313 (2015).

[40] L. Novotny and B. Hecht, *Principles of Nano-optics* (Cambridge University Press, Cambridge, 2012).

[41] L. X. Ge, Y. P. Cang, K. Gong, L. H. Zhou, D. Q. Yu, and Y. S. Luo, AIP Adv. **8**, 085321 (2018).

[42] *Handbook of Optical Constants of Solids*, edited by E. Palik (Academic, New York, 1998).

[43] E. Begüm Elçioğlu, A. Didari, T. Okutucu-Özyurt, and M. Pinar Mengüç, J. Phys. D: Appl. Phys. 52 105104 (2019).